# Motivational activation enables self-directed online search to affect policy support

Celina Kacperski,[1]† Roberto Ulloa,[1,2]† Peter Selb, [1,3] Andreas Spitz,[4] Denis Bonnay,[5] Juhi Kulshrestha[1,6]

[1]Cluster of Excellence "The Politics of Inequality", Konstanz University
[2]Humboldt University, Berlin
[3]Department of Politics and Public Administration, University of Konstanz
[4]Department of Computer and Information Science, University of Konstanz
[5]Department of Philosophy, Université Paris Nanterre
[6]Department of Computer Science, Aalto University

Corresponding author: roberto.ur@protonmail.com
†These authors contributed equally to this work.

## Statements and Declarations

### Acknowledgments

We owe many thanks to Katharina Jäger and Corinna Nitsch for their help with data annotation, and Julian Schelb and Anton Pogrebnjak for their support in data collection.

### Funding

Deutsche Forschungsgemeinschaft (DFG – German Research Foundation) under Germany's Excellence Strategy – EXC-2035/1 – 390681379

### Author contributions

Conceptualization: CK, RU, JK, PS, AS, DB
Data Curation: RU, CK, DB
Methodology: CK, RU, PS
Investigation: CK, RU
Data analysis: CK, RU, PS
Visualization: CK, RU
Funding acquisition: JK, PS, AS, DB
Writing—original draft: CK, RU
Writing—review & editing: CK, RU, PS, JK, AS, DB

### Competing interests

Authors declare that they have no competing interests.

### Ethics Approval

Ethics approval was received by the IRB under the number IRB23KN02-003/w, with consent collected from participants at all contact points

**Materials Availability Statement**
Complete materials are available in the supplemental materials.

**Data Availability Statement**
Survey and aggregated browsing data to reproduce the findings of this study is available at: https://github.com/robertour/s2j_attitude_change. Detailed behavioral browsing data (containing information that could compromise the privacy of research participants) can be made available to researchers upon reasonable request. Analysis scripts are available at: https://github.com/robertour/s2j_attitude_change

**Transparency Statement**
A preregistration was recorded on OSF ahead of rollout of the study. The present analysis deviates as follows. First, the primary outcome was changed from a discrete choice modelling task to Likert-scale policy support items. The choice modelling task could not be used due to a randomization implementation error by the survey/panel provider. The Likert-based attitude items, which were included in the survey alongside the choice task, are therefore used as the primary outcome here. Second, the primary analysis strategy was changed from linear mixed-effects regression to OLS for ITT effects and two-stage least squares instrumental variable regression for LATE estimates, to better address endogeneity in the relationship between information exposure and attitude change. All other design elements are consistent with the preregistration.

**AI usage statement**
This manuscript made use of Claude to assist with language polishing and consistency, and writing clarity improvements. The authors reviewed and edited all outputs, and they take full responsibility for the final text.

**Abstract**

As citizens increasingly encounter political information in digital environments, understanding whether this engagement shapes their policy views has become a central concern. Drawing on dual-process theories of persuasion, we argue that motivational activation is an enabling condition for policy support change in high-choice online environments. We test this in a three-wave field experiment with German participants (n = 791) across three policy topics (basic child support, renewable energy transition, cannabis legalization), in which participants were randomly assigned to a control group, and two encouragement conditions: a verbal encouragement, or a monetary incentive tied to a knowledge test. Browsing behavior was passively tracked via digital trace data over a 20-hour window. We find that self-directed online information search produced changes in policy support for child support and cannabis legalization but not for the energy transition, with monetary incentives producing significant effects rather than verbal prompts. We discuss motivational salience, issue malleability, and search-environment quality as joint conditions under which political information engagement can produce detectable changes in policy support.

Keywords: motivated reasoning, online information search, high-choice media environments, field experiment, digital trace data, policy support

**Materials Availability Statement**
Complete materials are available in the supplemental materials.

**Data Availability Statement**
Analysis scripts, survey and aggregated browsing data to reproduce the findings of this study are available at [redacted, will be made available upon revision]. Detailed behavioral browsing data (containing information that could compromise the privacy of research participants) can be made available to researchers upon reasonable request.

**Transparency Statement**
A preregistration was recorded on OSF ahead of rollout of the study. The present analysis deviates as follows. First, the primary outcome was changed from a discrete choice modelling task to Likert-scale policy support items. The choice modelling task could not be used due to a randomization implementation error by the survey/panel provider. The Likert-based attitude items, which were included in the survey alongside the choice task, are therefore used as the primary outcome here. Second, the primary analysis strategy was changed from linear mixed-effects regression to OLS for ITT effects and two-stage least squares instrumental variable regression for LATE estimates, to better address endogeneity in the relationship between information exposure and attitude change. All other design elements are consistent with the preregistration.

**LLM usage statement**
This manuscript made use of LLMs to assist with language polishing and consistency, and writing clarity improvements. The authors reviewed and edited all outputs, and they take full responsibility for the final text.

**Motivational activation enables self-directed online search to affect policy support**

## Introduction

In contemporary democracies, citizens form political preferences and make voting decisions in environments saturated with digital information. The move from broadcast-era media to high-choice, online ecosystems has reshaped how people encounter and evaluate information about public policy (Newman et al., 2025; Van Aelst et al., 2017). Ideally, this transformation holds democratic promise: when citizens have access to diverse sources, they can form attitudes that better align with their interests and values, improve their understanding of complex issues, and hold political actors accountable (Aalberg & Curran, 2013).

Yet despite this potential, recent political communication research has shown that exposure to political information produces only modest attitude change (Bennett & Iyengar, 2008; Coppock et al., 2020; Kubin & Von Sikorski, 2021). It has been argued that even in contemporary high-choice online environments, where citizens encounter diverse, policy-relevant information across news sites, social media, and search engines, prior attitudes, partisan identities, and cognitive filters constrain the influence of consumed content (Guess et al., 2021; Wojcieszak et al., 2023). Recent large-scale field experiments on social media platforms reinforce this picture: sustained algorithmic interventions lasting several months produced limited effects on political attitudes and polarization (Guess et al., 2023; Nyhan et al., 2023). This raises the question of under what conditions political information engagement does produce attitude change, a question this study seeks to address.

A growing body of work argues that effect sizes and their variability may reflect how citizens encounter political information in experimental settings, that is, the nature of information exposure (Arceneaux et al., 2013). Much of today's political learning takes place in contexts where individuals exercise substantial control over what information to access, when, and through which sources. Theorizing on dual-process persuasion and motivated reasoning suggests that motivational processes, particularly the decision to seek out information at all, are central to whether and how political attitudes are formed or revised (Kunda, 1990; Stroud et al., 2019). If individuals are insufficiently motivated to attend to, or evaluate incoming content, information processing may fail to initiate, not because the messages are ineffective per se, but because they are never deeply engaged with. In addition, experiments often rely on researcher-controlled exposure: participants are shown specific articles, videos, or arguments and then asked to evaluate them. These designs afford high internal validity, isolating message features under well-defined conditions (Amsalem & Zoizner, 2022), but this methodological strength may come at a cost. Experimental evidence on political attitude change under conditions of motivated, self-directed online search remains rare, despite its potentially central relevance in high-choice media environments where the most recent field experiments have been conducted (Aslett et al., 2024; Casas et al., 2023; Wojcieszak et al., 2022).

In this study, we seek to advance the existing literature on political attitude change by combining motivational activation with agency over information acquisition in an experimental setting. Specifically, across three waves of data collection, we embed participants in an unconstrained online environment in which they are either implicitly or explicitly encouraged to search for information on three preset topical policy issues (child support, renewable energy transition, cannabis legalization). We vary the form of motivational activation in the explicit-encouragement condition, providing either a verbal accuracy prompt or a monetary incentive tied to a knowledge test. We passively observe participants' browsing behavior using digital trace data. This design allows us to examine whether individuals change their political attitudes following verified information search.

Our aim is to build on prior literature on the effects of media on attitudes by testing the preconditions under which change might occur. If, as dual-process and motivated-reasoning theories suggest, policy attitude change depends on the willingness to process information, then a design that allows participant agency to search the internet freely and activates motivation more strongly via explicit encouragement should help identify conditions under which attitude change becomes detectable. In doing so, this study offers new insight into the motivational, behavioral, and issue-level conditions that shape the degree to which political information engagement translates into attitude revision.

## Theoretical Background

### Cognitive models of attitude updating

Dual-process models of persuasion such as the Elaboration Likelihood Model (ELM) (Petty & Cacioppo, 1986) and the Heuristic-Systematic Model (HSM) (Chaiken, 1980) offer a foundational basis for understanding political attitude change, by disdistinguishing between deeper, effortful processing and superficial processing. These are elaborated by work on epistemic motivation (Kruglanski, 1990) which emphasizes that people's approach to knowledge acquisition (with how much urgency and enjoyment they do it) matters when forming or updating attitudes (Arceneaux & Vander Wielen, 2013); and by work on motivated reasoning (Kunda, 1990), where accuracy motivations and directional goals have been shown to enhance information integration and reduce partisan bias (Christensen & Moynihan, 2024; Diana et al., 2020; Druckman, 2012). It is useful to distinguish the theoretical roles these constructs play in online political attitude formation (Arceneaux et al., 2025): epistemic motivation and intrinsic interest govern whether and how energetically someone initiates information search; ELM/HSM elaborate whether, once encountered, that information is processed systematically or heuristically; and motivated reasoning specifies the directional versus accuracy goals that shape how evidence is integrated and which conclusions are tolerated. These constructs are related, and a central implication is that motivational states operate at multiple stages of attitude revision.

Recent experimental work has begun to research in this direction, with consequences for the long-running motivated-reasoning debate. The motivated system 2 reasoning account first held that more deliberation should amplify partisan bias by giving individuals greater capacity to rationalize prior commitments (Kahan, 2013, 2016), but then, a growing experimental literature challenged this view. Pennycook and Rand (2019) found that analytic thinking predicted better discernment of true from false news regardless of partisan congruence. Bago et al., (2020) showed that deliberation reduced belief in false headlines without inflating polarization. Tappin et al. (2020, 2021) further demonstrated that cognitively sophisticated individuals appear to update more responsively to new information, and finally, no association was found between numerical ability and politically motivated reasoning in a large nationally representative U.S. sample (Stagnaro et al., 2023). Together, these studies suggest that engaging analytic processing is usually accuracy-promoting rather than amplifying bias, in line with the original premise of dual-process accounts (Ditto et al., 2025).

If deliberation, when activated, generally moves people toward accuracy and/or attitude-choice alignment, the question is then how to trigger deeper processing. Tappin et al. (2020) make the methodological point here that standard experimental designs in this field traditionally have randomly assigned information, not motivation. To probe whether motivation actually constrains attitude revision, an experimental design must manipulate motivational state directly and, ideally, observe what individuals do with the information environment when given the freedom to engage with it on their own terms.

**From accuracy goals to monetary stakes**

In line with the idea that motivation is the binding constraint, Pennycook et al. (2021) showed across multiple experiments that accuracy prompts increased the quality of news participants intended to share. A meta-analysis of twenty experiments (N = 26,863) found that accuracy prompts reduced sharing of false headlines by roughly 10% relative to control without significant decay or ideological asymmetry (Pennycook & Rand, 2022) and a 16-country, 34,286-participant cross-cultural replication confirmed that these effects generalize well beyond U.S. samples (Arechar et al., 2023).

There is also evidence that stronger motivational manipulations produce correspondingly stronger effects, such that monetary incentives improved accuracy and reduced partisan bias in headline judgments by approximately 30% (Rathje et al., 2023). This research extends an earlier line of work on partisan motivated answering, where monetary incentives and accuracy appeals jointly reduced partisan gaps in economic perceptions (Prior et al., 2015). Finally, in a field experiment, Bowles et al., (2025) found that financial incentives increased actual consumption of fact-checked content and improved misinformation discernment in subsequent assessments, a behavioral analogue to previous survey-based findings. Taken together, the evidence suggests that explicit accuracy motivations work primarily by recruiting attention and effort to information that participants would otherwise process superficially, with monetary stakes outperforming verbal prompts when the underlying task involves actively engaging with information rather than merely reporting beliefs. This literature focuses primarily on misinformation detection and news sharing rather than policy attitude change, so the transferability of these findings is not automatic, however, the underlying mechanism, that motivational activation increases the depth and breadth of information engagement, which in turn shapes evaluative outcomes, we expect to be general enough to apply across outcome domains.

**Self-directed search in high-choice environments**

More recently, researchers have begun to study political attitude change in genuinely high-choice environments, often using digital trace data to observe how participants engage with online information when given broader instructions. Yet on average, such studies are more likely to report null or mixed effects on political attitudes. Casas et al., (2023) found that incentivized visits to opposing partisan news sites produced limited effects on attitudes. Wojcieszak et al. (2023) using over twelve months of web-browsing data, found that real-world exposure to liberal, conservative, or centrist news produced little or no causal effect on affective or attitude polarization. Aslett et al. (2022) demonstrated no effects of adding source credibility labels on participants' affective polarization and political cynicism. Guess (2021, 2023) did not find significant effects in two experiments on their attitude and opinion measures, though exposure to either left or right partisan media reduced trust and confidence in mainstream media. Levy (2021) manipulated Facebook news feeds and also found no evidence of effects on political opinion, but observed reduced negative attitudes towards the opposing political party. Only Allcott et al. (2020) found that deactivation of Facebook reduced both factual political knowledge and self-reported issue polarization, though not affective polarization, and Bail et al. (2018), paying participants to follow opposing-view Twitter bots, did find that Republicans became substantially more conservative.

Several recurring features of these designs may help explain the predominance of null or small effects. First, they typically improve ecological validity by simulating real-world media behavior, but even when participants are nominally in a high-choice environment, the proportion who actually consume political information is reportedly small (Mangold et al., 2024; Wojcieszak et al., 2022), so the effect of offering a richer information environment might differ from the local effect among compliers (those who actually engage). Also, even compliance is not a guarantee of attitude-relevant or high-quality informational outcomes: Aslett et al., (2024) showed across five experiments that searching online to evaluate misinformation can increase the perceived veracity of false stories, an effect concentrated among individuals for whom search engines return lower-quality results from unreliable sources, and Robertson et al., (2023) found that engagement with unreliable news on Google Search was driven primarily by users' own partisan choices. The information environment encountered during self-directed search is therefore not neutral: it depends on who is searching, what they search for, and what the search architecture surfaces.

Finally, a common feature of these designs is the absence of motivational activation: participants control their viewed content without any particular reason to engage with it, but control over the choices may let participants pursue epistemic goals only if such goals are made salient. It is possible therefore that agency over information acquisition is not sufficient for attitude revision.

**Heterogeneity of policy issues and corresponding information environment**

A final theoretical consideration concerns heterogeneity across issues. Even when motivation and agency are jointly activated, the malleability of an attitude depends on (1) what participants already believe, for example Druckman & Leeper (2012) show that persuasion effects shrink as priors crystallize through previous exposure, a finding that updates Zaller's (1992) classic account that long-running, identity-laden issues are harder to move than recent ones; and (2) and on the structure of the information environment, i.e., how salient the topics are and how are they presented, for example Gramacho et al. (2026) shows substantial variation in effects due to information environment changes.

For the German policy context, the three policies relevant in the following study were selected in part to vary on theoretically relevant dimensions such as salience, complexity and divisiveness (as ratified via a pretest conducted before the study, described in detail in Supplement SM). On this basis, we form the exploratory expectations that attitudes on the three topics might shift differently: the renewable-energy transition (Energiewende) has been in the public eye for over more than two decades of political contestation, with attitudes structured by ideology, regional infrastructure conflict, and identity-relevant cultural meanings (Sonnberger et al., 2021; Sonnberger & Ruddat, 2017). German cannabis attitudes in contrast, have been shown to be tied to expectations about consequences, which might make them more sensitive to new information (Manthey et al., 2024). Finally, both basic child support policy and cannabis legalization were novel policy debates in parliament, with child support policy particularly less ideologically pre-loaded, with substantial uncertainty in the public making information more likely to register. The same set of motivational and agency conditions might therefore yield heterogeneous effect sizes depending on the malleability of the underlying issue.

**A motivation–agency account of attitude updating**

In summary, models of attitude change consistently identify motivation and effort as necessary conditions for belief revision; and these conditions interact with the issue context in ways that are consistent across communication modalities (Settle, 2025). Recent experimental work suggests that, when activated, deliberation tends to move attitudes toward accuracy rather than entrench priors; and high-choice digital trace studies suggest that agency without motivation tends to produce small, mixed, or null attitude effects on average. We test this account by jointly varying motivational activation (implicit, explicit verbal and explicit monetary encouragement) and allowing agency, i.e., self-directed search in an unconstrained environment.

Two predictions follow. First, H1, attitude change should be more likely to emerge under explicit motivational activation than under implicit-search conditions, with monetary incentives plausibly outperforming verbal prompts (Bowles et al., 2025; Rathje et al., 2023). Second, H2 (exploratorily) that the magnitude of attitude change may vary, with greater movement expected on issues where public uncertainty is higher and partisan crystallization is lower (Druckman & Leeper, 2012; Slothuus & Bisgaard, 2021).

Thus, the present study aims to clarify the boundary conditions under which citizens, given both the freedom and the reason to inform themselves, report changing their minds.

## Methods

### Survey design and data collection

Ethics approval was received by the IRB under the number IRB23KN02-003/w, with consent collected from participants at all contact points (see SK). The trial design is illustrated in Figure 1.

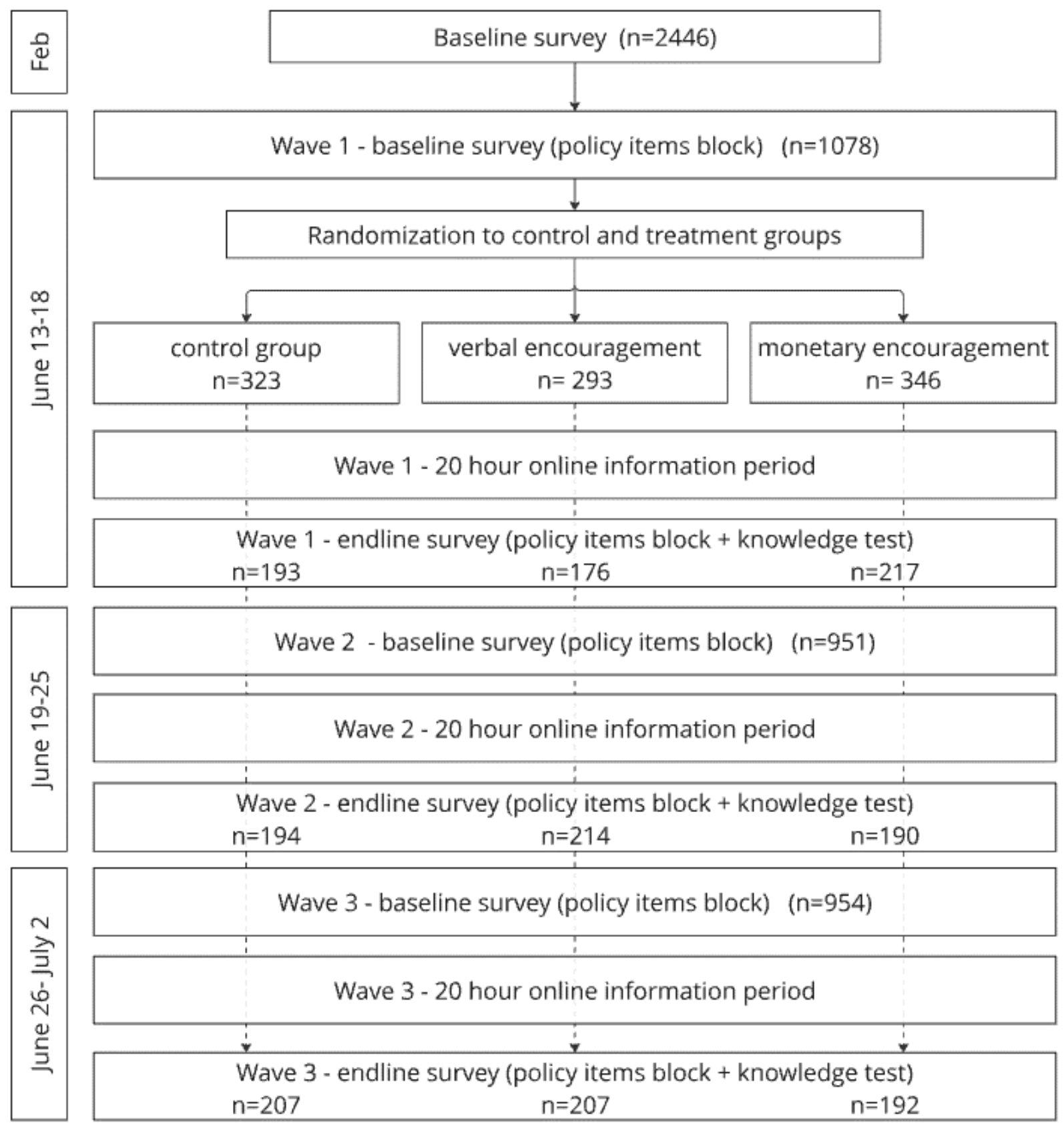

**Figure 1. Experimental design.** Participants were randomized at the end of the first baseline survey and remained in the same group thereafter. They received the intervention text according to their group assignment (see Table 1). They then had 20 hours to inform themselves on the internet. In the post-phase, they again received the same policy items and the knowledge test. This was repeated for all three waves.

First, all participants that were part of the web tracking panel received an invitation to join in a study in February 2023 that lasted 20 minutes and included a battery of survey items querying socio-demographics, political interests, knowledge and attitudes. In June 2023, all participants that had completed the February introductory survey were reinvited to the main experimental trial. The experiment ran in three experimental waves using three policies that had been selected due to being prevalent in the news as they were being discussed in the German parliament, and that had been pretested to vary on the dimensions of salience, divisiveness and complexity, while being rated as interesting (for more details on topic selection, see SM and Ulloa et al., 2026): (1) basic child support policy (Kindergrundsicherung); (2) renewable energy transition policy (Förderung erneuerbarer Energien) and (3) cannabis legalization policy (Cannabislegalisierung). Participants were informed that each survey would take about 15 minutes, run across two contact points one day apart, and in the baseline answered a series of questions about the policy topic (see SJ). At the end of the first contact point (baseline), participants were randomized into the three intervention groups and instructed to return to the survey 20 hours later to complete the postline survey on the following day. Participants in the control group were not given any additional instructions; participants in the verbal encouragement group were encouraged to inform themselves on the policy topic relevant in each respective wave; participants in the financial encouragement group were encouraged to inform themselves on the policy topic and that they could earn 3 Euros if they managed to pass a knowledge test at the end of the survey (3 out of 5 questions correctly answered). Knowledge findings are reported in a companion article (Ulloa, 2026). After 20 hours, a reminder email was sent to participants, who, if returned, completed the postline survey and knowledge test. Trial details, including exact wordings, can be found in SA.1.

Participants of our study consented to install tracking software on their desktop and/or mobile devices. The software records URL-level web browser views (a visited URL can be viewed multiple times as the tab can be activated several times), including access time and duration. During the 20-hour intervention periods, participants accumulated ~254K URL visits (M=170.22, Mdn=103, SD=204.47), corresponding to ~209K quasi-unique URLs (i.e., the sum of the total unique URLs per topic), which were manually annotated by two research assistants according to whether (or not) the URLs from each wave were directly related to their relevant policy topics. For further details on the annotation process, see SH.1.

**Participant details**

Of the n=2446 participants in the February survey, n=1112 German-speaking residents of Germany were recruited to our experiment via a panel provider across all three waves, and compensated at the provider's usual rate (6€/hour), making our sample a convenience sample. n=315 were excluded due to missing web activity logs, and n=6 due to a failed attention test (see details in SA.2), with the final sample being n=791 participants, with slight variations across the three data collection waves, as participants were allowed to join in later waves even if they did not participate in a previous wave ($N_{wave1}$=586, $N_{wave2}$=598, $N_{wave3}$=606). Exclusions did not differ between the intervention groups in any of the waves, neither for technical exclusions ($p>0.35$), nor attention check exclusions ($p>0.19$) (see SA.5 for more details). Of the final sample, 363 (45.9%) identified as women, 428 (54.1%) identified as men (the German population skews 51.7% female), so our sample includes more men than the general population. No participants identified as other gender categories. Participants ranged in age from 18 to 78, with a median age of 50 (mean=48.6, SD=12.31 – population median is 44.7, so our sample is somewhat older than the general population). Educational attainments were summarized into low (elementary school or equivalent, n=98, 12.4%), medium (middle school/vocational certificate or equivalent, n= 328, 41.5%), high (high school certificate or equivalent, n=155, 19.6%), and very high (university degree or equivalent, n=210, 26.5%); the German population overall reports less vocational level educated individuals (up to middle school, 47.5%) and less higher education degrees (17.6%), and more individuals that finished higher education entrance qualifications (high school eq., 31.9%) (Statistisches Bundesamt, 2024).

**Analysis**

Our design identifies intent-to-treat (ITT) effects and, under standard assumptions, the local average treatment effect (LATE; Angrist & Imbens, 1995), for which we used instrumental variable (IV) regression. Following standard assumptions (Sovey & Green, 2011), random encouragement to seek information serves as a valid instrument for actual information search: it is exogenous by design and predicts online information consumption. The exclusion restriction requires that encouragement affects attitude change only through verified online information engagement, and not through alternative pathways. We discuss the plausibility and limits of this assumption in the Limitations section, as well as details and a thorough discussion of all assumptions are provided in SA.4.

IV models were estimated via two-stage least squares using *ivreg* in R. Analyses using binary instruments are reported for each verbal and monetary condition vs control, with further details in SC. ITT effects were estimated using OLS with intervention condition as a factor. Outcomes were absolute and directed attitude change scores (post–pre). The main endogenous predictor, exposure to policy-related online content, was coded as a binary indicator of at least one valid visit, a standard practice in IV design where take-up is sparse, as this maximizes statistical power in 2SLS estimation (Sovey & Green, 2011). Additionally, alternative operationalizations (counts, search queries) and models with covariates yield substantively identical results and are reported in SC and SD.

## Results

### Descriptive results

In Table 2 we report descriptives for policy-relevant content visited (articles viewed and searches conducted), as well as time spent on engaging with the visits and searches. The values suggest that the encouragements, particularly the monetary incentive, increased both searches and URL visits. There is also an increase for the intervention groups for time spent on URLs and searches. We also report means and standard deviations of attitude change and directionality variables. Here, values are higher for the intervention groups in the basic child support wave, as well as the cannabis legalization wave. Distribution of participants across latent groups of compliers, always-takers and never-takers and can be found in SA.3.

**Table 2. Descriptives of predictor and response variables across the three policy waves.** Absolute numbers given, percentages in brackets.

| | **Child support** | | | **Energy transition** | | | **Cannabis legalization** | | |
|---|---|---|---|---|---|---|---|---|---|
| **Intervention group** | **control n=193** | **verbal n=176** | **monetary n=217** | **control n=194** | **verbal n=214** | **monetary n=190** | **control n=207** | **verbal n=207** | **monetary n=192** |
| Participants with policy-related URL visits | 3.11% | 21.0% | 37.3% | 3.09% | 15.0% | 27.4% | 2.90% | 17.4% | 25.0% |
| Avg policy-related visits | 0.04 (0.25) | 0.30 (0.65) | 0.88 (1.54) | 0.04 (0.21) | 0.38 (1.31) | 0.86 (2.70) | 0.04 (0.26) | 0.26 (0.65) | 0.53 (1.52) |
| Avg policy-related URL visits n(visits)>0 | 1.33 (0.52) | 1.43 (0.65) | 2.36 (1.69) | 1.17 (0.41) | 2.56 (2.46) | 3.15 (4.43) | 1.33 (0.82) | 1.50 (0.74) | 2.10 (2.45) |
| Participants with policy-related searches | 5.18% | 22.2% | 38.2% | 4.12% | 16.4% | 24.2% | 4.35% | 19.8% | 27.6% |
| Avg policy-related searches | 0.09 (0.43) | 0.33 (0.73) | 0.93 (1.86) | 0.12 (0.86) | 0.34 (1.03) | 0.64 (1.65) | 0.04 (0.20) | 0.31 (0.77) | 0.56 (1.18) |
| Avg policy-related searches n(searches)>0 | 1.80 (0.79) | 1.49 (0.82) | 2.43 (2.32) | 3.00 (3.25) | 2.06 (1.73) | 2.65 (2.45) | 1.00 (0.00) | 1.59 (0.97) | 2.04 (1.44) |
| Avg time spent on policy-related URL visits (seconds) | 88.3 (74.3) | 143 (125) | 337 (343) | 41.7 (23.2) | 170 (194) | 281 (476) | 159 (172) | 238 (265) | 420 (724) |
| Avg time spent on policy-related searches (seconds) | 63.1 (51.4) | 79.5 (155) | 84.4 (120) | 59.8 (58.4) | 82.2 (92.4) | 106 (156) | 34.7 (33.8) | 48.2 (55.4) | 53.2 (57.2) |
| Attitude towards policy topic (pre) | 4.99 (1.77) | 4.84 (1.66) | 4.87 (1.76) | 5.29 (1.63) | 5.27 (1.70) | 5.34 (1.50) | 3.92 (2.10) | 4.01 (2.20) | 4.20 (2.15) |
| Attitude towards policy topic (post) | 4.92 (1.78) | 5.12 (1.75) | 5.08 (1.74) | 5.17 (1.59) | 5.21 (1.73) | 5.35 (1.54) | 4.00 (2.13) | 4.00 (2.30) | 4.16 (2.14) |
| Attitude (absolute change) | 0.65 (0.94) | 0.78 (1.02) | 0.93 (1.15) | 0.76 (1.00) | 0.70 (0.98) | 0.78 (0.99) | 0.44 (0.75) | 0.60 (1.08) | 0.69 (1.09) |
| Attitude (directed change) | -0.07 (1.14) | 0.28 (1.26) | 0.21 (1.46) | -0.12 (1.25) | -0.07 (1.20) | 0.01 (1.26) | 0.08 (0.86) | -0.01 (1.23) | -0.04 (1.29) |

## Changes in attitudes

We examine the intention to treat (ITT) effects, see upper half in Table 3. For this, we employ linear regression models, predicting attitude change from the interventions (verbal and monetary encouragement) compared to the control group. We present the two interventions separately to showcase that the addition of a monetary incentive to the encouragement affects attitude change significantly, more so than the verbal instruction alone. Effects are significant for the two policy topics child support and cannabis legalization, but not for the energy transition policy (see SB, Figure S3 for a plot of means across intervention groups and policy topic waves).

**Table 3. Attitude changes as a result of intervention and online search treatment.** Upper half: Results from an OLS linear regression estimating intention to treat effect (ITT) of interventions (verbal and monetary encouragement) on attitude change. Lower half: Results from 2SLS instrumental variable regression predicting attitude change from tracked visits to policy-relevant URLs, averaged across the two treatment groups. Individual analyses for verbal and monetary are in SC. FDR adjusted p-values (method BH) are reported in brackets behind significant p-values (Benjamini & Hochberg, 1995))

| **OLS linear regression - Intention to treat effect on attitude change** | | | |
|---|---|---|---|
| | ***Child support*** | ***Energy transition*** | ***Cannabis legalization*** |
| **Verbal encouragement** | **0.057, p=0.228** | **-0.028, p=0.563** | **0.077, p=0.099** |
| Confidence intervals | [-0.036, 0.151] | [-0.121, 0.066] | [-0.014, 0.168] |
| **Monetary encouragement** | **0.128, p=0.007 (0.018) **** | **0.010, p=0.834** | **0.117, p=0.012 (0.018) *** |
| Confidence intervals | [0.035, 0.222] | [-0.084, 0.104] | [0.026, 0.209] |
| R2 | 0.012 | 0.001 | 0.011 |
| F | 3.636 | 0.339 | 3.296 |
| **2SLS instrumental variable regression – Local average treatment effect on attitude change** | | | |
| | ***Child support*** | ***Energy transition*** | ***Cannabis legalization*** |
| | | Verbal encouragement treatment | |
| **Tracked visits to policy-relevant URLs** | **0.240 p=0.212** | **-0.141 p=0.565** | **0.357 p=0.108** |
| Confidence intervals | [-0.137, 0.618] | [-0.621, 0.339] | [-0.079, 0.794] |
| Weak IV F-stat (p) | 30.9 (0.000) | 17.6 (0.000) | 25.2 (0.000) |
| Wu-Hausman (p) | 1.635 (0.202) | 0.150 (0.698) | 1.61 (0.205) |
| Num.Obs. | 369 | 408 | 414 |
| | | Monetary encouragement treatment | |
| **Tracked visits to policy-relevant URLs** | **0.313* p=0.012** | **0.032 p=0.835** | **0.412* p=0.014** |
| Confidence intervals | [0.069, 0.557] | [-0.266, 0.329] | [0.083, 0.741] |
| Weak IV F-stat (p) | 86.3 (0.000) | 49.6 (0.000) | 46.2 (0.000) |
| Wu-Hausman (p) | 8.486 (0.004) | 0.125 (0.724) | 7.416 (0.005) |
| Num.Obs. | 410 | 384 | 399 |

For the IV regression, we expected that participants who visited topic-relevant webpages (compared to those who do not) would on average report stronger changes in policy attitudes. Table 3 (lower part) presents the results of the 2SLS instrumental variable regression[1]: for the monetary encouragement group, if a participant visited at least one article on the given topic (as predicted by the group), their reported attitude changed by approximately 0.31 points for child support and 0.41 points for cannabis legalization, while no significant effect was observed in the energy transition policy wave. For the verbal encouragement, we did not find any significant differences to control, though the effect sizes for child support and cannabis legalization policies were found to be in a consistent direction.

[1] Due to the two-stage process, the standard errors in 2SLS are generally larger than those in OLS because there is additional estimation error introduced in the first stage (when estimating the predicted values of the endogenous variables using the instruments). Larger standard errors typically make it harder to achieve statistical significance, making 2SLS more conservative in this sense compared to OLS. This speaks to the robustness of our results.

We explored the directionality of the attitude shifts (see SF, for illustrative purposes, tile plots in Figure S4 in SF), and we observe a sample-wide average directional shift in the positive direction for the treated for child support, while we observe a pattern of attitude changes without any specific direction across the sample for cannabis legalization. Models here show significant effect of online search on directed attitude change for child support policy in the observed direction, $b=0.258$, $p=0.034$, CI95%[0.020, 0.497], while indeed we do not find any significant directed effect neither for cannabis legalization, $b= -0.178$ $p=0.271$, CI95%[ [-0.143, 0.451]), nor for energy transition policy, $b= 0.154$ $p=0.309$, CI95%[-0.497, 0.140]. Additionally, visually, no pattern of attitudinal reinforcement or polarization is discernable, and we did not find any significant ITT or LATE effects when modelling this with linear regressions, i.e., participants were not more likely to shift towards a more extreme pre-existing position. ($p$’s $> 0.374$).

Based on our theorizing regarding heterogeneity effects, and pretest ratings of policy topics across various dimensions (SM), we formed an exploratory hypothesis that topics might have differed in divisiveness across party lines. To check this, we aligned attitude ratings with party affiliation of participants. Descriptively, as can be seen in the density graph in Figure 2, it appears that the energy transition policy split attitudes across party lines the strongest, with much higher opposition and lower support from far-right-wing parties (blue line) and very strong support from left parties (red & green lines). Comparatively, cannabis seems generally divisive, with individuals from most parties both opposing and supporting, i.e. less so in line with party affiliation.

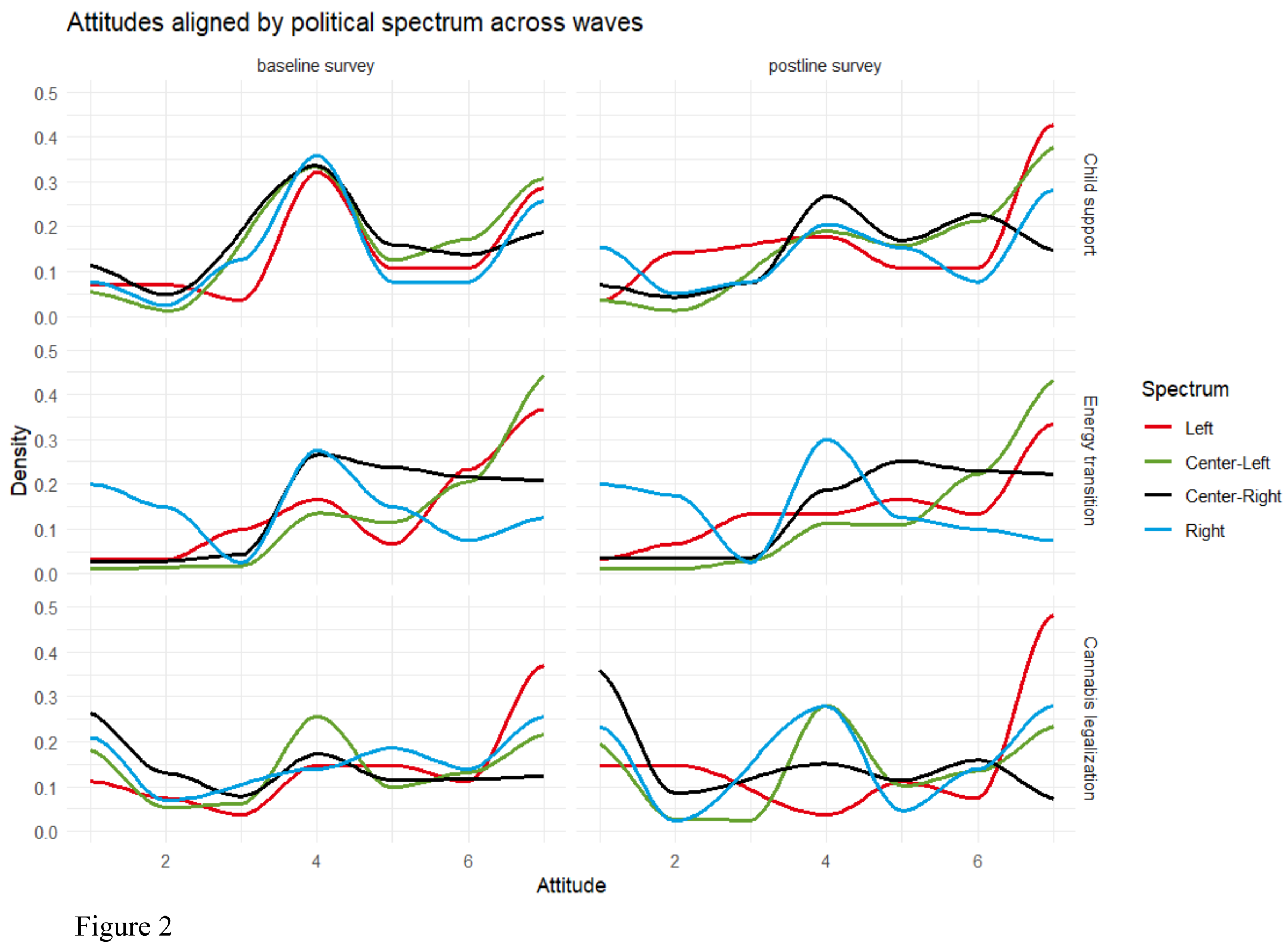

Figure 2

**Figure 2. Attitudes of treated participants based on party affiliation.** Party affiliation is coded on a left-to-right spectrum: Left party (far left), Green party and SPD (center-left), CDU/CSU and FDP (center-right) and AfD (far right). The plots only include data of the verbal and monetary treatment groups to compare between pre-treatment baseline (left) and post-treatment postline (right).

We also queried participants' attitudes towards specific policy-related issues, expecting that we would find changes as a result of online information engagement. We present the results of 2SLS instrumental variable regressions of the effect of online search on attitude changes regarding these specific issues in Figure 3. Tables of all 2SLS and OLS linear regressions are reported in SE. We find three significant effects in the child support policy for absolute change, and four significant effects in the directed change in a positive direction. We also find a significant effect of online search on the oil/gas heater ban, and one significant negative effect regarding the purchase limit of cannabis. While the child support policy related effects remain significant at FDR correction, the other two effects are not significant when corrected (see SE for all corrected p-values).

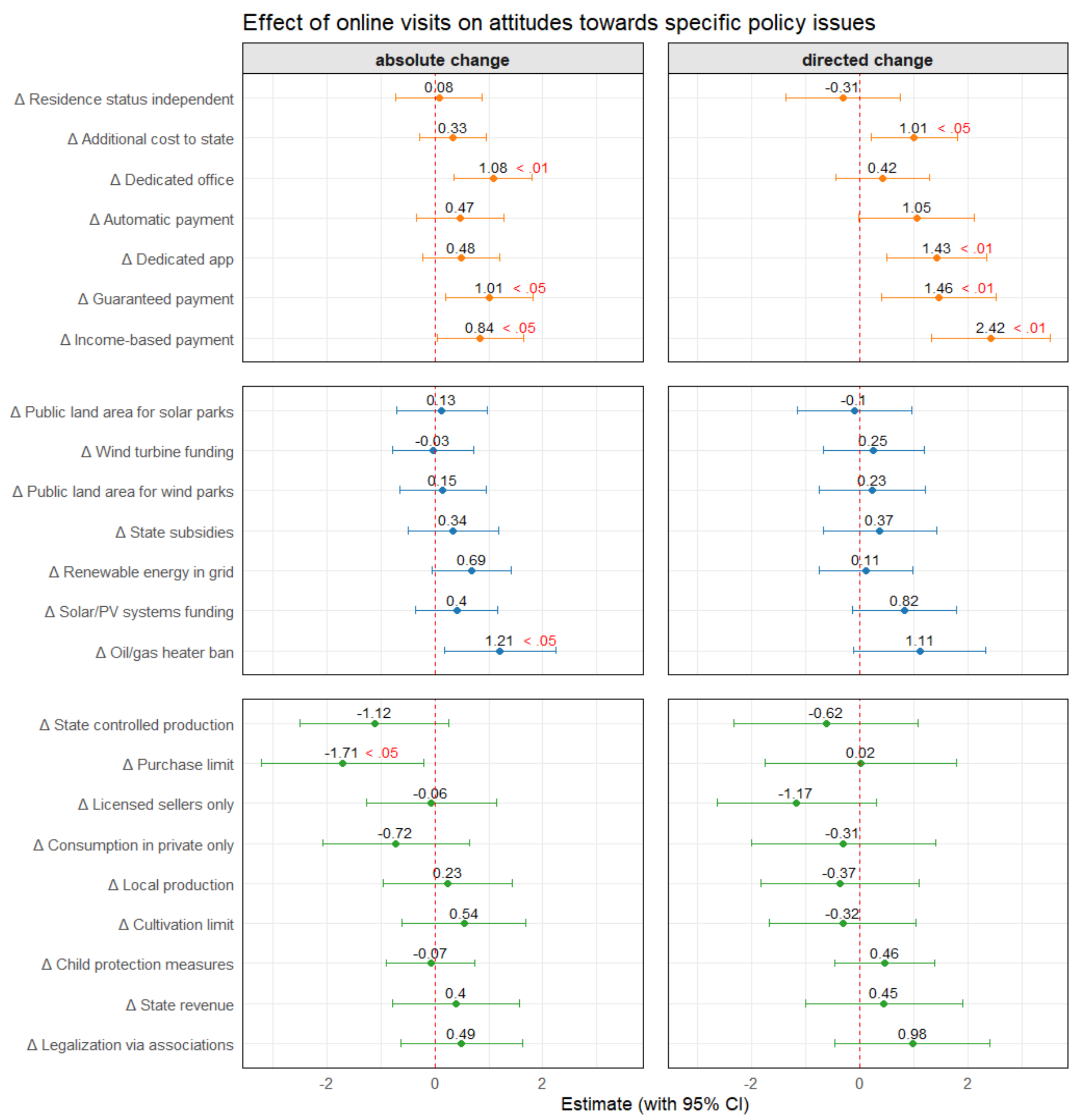

Figure 3. **2SLS instrumental variable effects of online search on attitudes towards specific policy-related issues.** Significant effects are labelled at p<0.05 and p<0.01 level and uncorrected. FDR corrected p-values

are reported in supplement SE, with most remaining significant, while heater ban and purchase limit $p$'s > .10.

## Information search in a high-choice environment

We provide elaboration about participants engaging in information search in a high-choice environment to illustrate the challenge of simulating the assignment of information to participants in a controlled set-up. Figure 4 showcases the patterns of policy-related information exposure triggered by our intervention in each of the topics. We observe numerous instances of websites that are visited by a single participant, indicating that those individuals received an exclusive treatment. At the same time, one URL in the cannabis legalization topic was selected by 41% participants (N=41), statistically higher than the most selected URL for child support (26.56%, N=34; $z = 2.72$, $p$-adj = .020, proportional z-test, p-values adjusted via Bonferroni correction) and for energy transition (26.53%, N=26; $z = 2.91$, $p$-adj = .014). Some participants visited many URLs, the ones with most visits did so for 22.02% (N=24) of the total visited URLs for the energy transition, 25.49% (N=13) for cannabis legalization and 8 URLs (11%) for child support. See SG.1 for distributions.

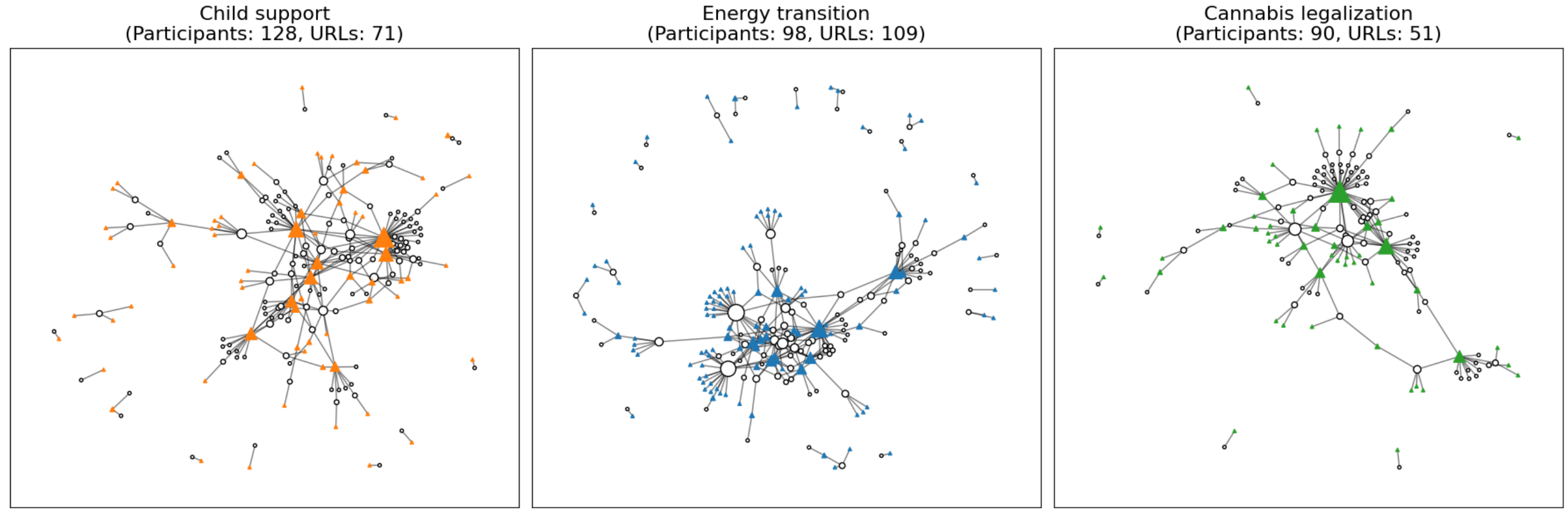

Figure 4. **Networks of participants and URLs.** Three bipartite graphs representing the visits of a participant to a URL, i.e., edges. Nodes are either participants (white circles) or URLs (triangles colored according to the policy).

We also observe differences between topics according to the type of URLs visited by participants, see Figure 5. For example, very few participants visited news sources when informing themselves about the energy transition policy (n=13), compared to when they were engaging with the child support policy (n=81) and cannabis legalization (n=56). The differences between the actual URLs (n= 36, 26, and 31 respectively) and domains (n=19, 16, 18) is not nearly as pronounced suggesting that although news covered the energy transition policy, participants in that group seldom selected them. Instead, more participants selected government pages (n=79) compared to child support (n=38) and cannabis legalization (n=46).

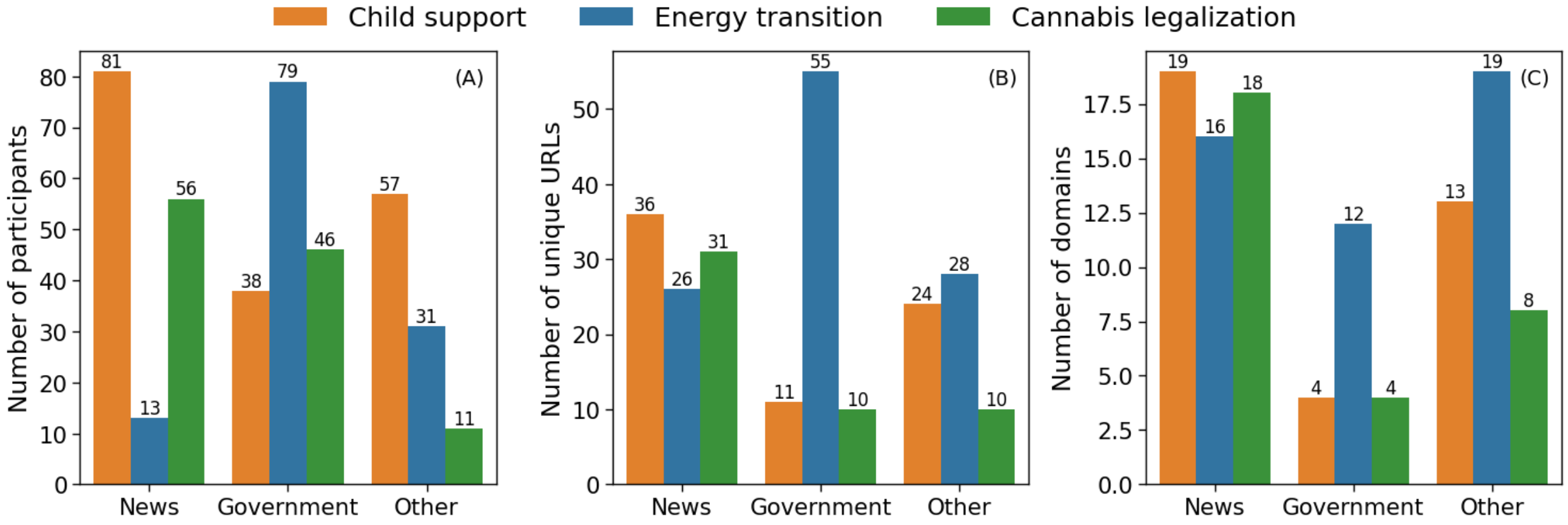

Figure 5. **Distribution of participants, domains and URLs according to type of pages.** The left plot (A) shows the number of participants (Y-axis) that had at least one URL associated with the respective type of content per policy topic (legend). The center plot (B) shows number of unique domains; the right plot (C) shows number of URLs that the participants encountered. X-axis splits by domain type.

Although we did not give any specific instructions to participants on how they should seek online information, most of the participants that choose to search (i.e., compliers and always-takers) limited themselves to a small number of searches using a commercial search engine (M=1.41, Mdn=1, SD=1.31). Google concentrated 83.83% of the searches. Comparatively, content visits averaged M=1.72 (Mdn=1, SD=2.23) per policy (see Figure S 10 in SG.3). We identified the referrers of the content visits, i.e., the visits preceding the topic-related (content) visit (see Figure S 11 in SG.3). The largest percentage of referrers were web searches: 49.84% for child support, 29.45% for energy transition, and 50.97% for cannabis legalization. The second category was topic-related visits of the same domain (direct referrals):

11.97% for child support, 19.74% for energy transition, and 11.17% for cannabis legalization.

## Discussion

We conducted a three-wave field experiment encouraging participants to inform themselves online about three German policies, basic child support policy, renewable energy transition policy, and cannabis legalization. We tested whether self-directed information engagement explicitly supported by either a verbal accuracy prompt or a monetary incentive tied to a knowledge test, would shift policy attitudes. The design was motivated by a theoretical position developed earlier in the paper: that motivation and agency may jointly increase the likelihood of belief revision, and that the average prevalence of small or null effects in high-choice digital trace studies (e.g., Guess et al., 2023; Nyhan et al., 2023; Wojcieszak et al., 2023) may reflect, at least in part, the absence of explicit motivational activation.

Our findings identify two conditions in high choice environments under which modest attitude change becomes detectable, motivational strength and issue malleability. Monetary encouragement consistently increased engagement with policy-relevant content: more searches, more URL visits, and longer time on page in the treatment groups than in the control group. Among compliers, this engagement translated into attitude change for child support (LATE ≈ 0.31 on a seven-point scale) and cannabis legalization (LATE ≈ 0.41), with directional shifts toward greater support for child support but no clear directional pattern for cannabis. We found no significant change for the energy transition policy topic. We also in exploration of the data did not find any evidence of attitude reinforcement or polarization in either ITT or LATE specifications, suggesting that, when search was self-directed and motivation made salient, participants did not retreat to more extreme prior positions. We did not find significant attitude change for any of the policies for the verbal encouragement group.

The estimated effects are modest (around 0.3–0.4 points on a seven-point Likert scale), but meaningful given the short duration of the experiment, as well as relative to the benchmark established by recent persuasion research, where for example an aggregated report of 59 randomized political-advertising experiments reported persuasion effects so small and uniform that no contextual moderator changed them appreciable (Coppock et al., 2020). Additionally, it has been repeatedly shown that even small individual-level shifts can be meaningful at the population level (Götz et al., 2022; Greenwald et al., 2015).

**Motivational activation as binding constraint**

While the comparison to control tells most about differences in initial activation to start an information search, the difference in significant effects of the verbal and monetary encouragements is theoretically informative regarding motivation: both interventions told participants to inform themselves about the policy, but only the monetary version added a reason to do so thoroughly. It then produced larger increases in observable search and browsing behavior (more searches, more page visits, more time on page) and stronger ITT and LATE effects on attitude change. We interpret this pattern as consistent with the dual-process accounts of persuasion (Chaiken, 1980; Petty & Cacioppo, 1986), which would predict this effect if motivation, rather than information availability, is the binding constraint on attitude change. We acknowledge at the same time that behavioral indicators such as visit counts and dwell time index the quantity of information engagement rather than its cognitive quality, and are thus imperfect proxies for cognitive elaboration; we return to this as a limitation below. Our findings are in line with recent evidence from the misinformation literature, where financial incentives have increased consumption of corrective content and subsequent discernment of falsehoods even in partisan contexts (Bowles et al., 2025; Prior et al., 2015; Rathje et al., 2023). In our experiment, this happens without any researcher selection of content. As the financial incentive is tied to whether participants pass a knowledge test after a self-directed search session, the primary behavioral channel we identify is information acquisition and engagement, with the observed attitude shift a downstream correlate of verified browsing among compliers. We cannot fully exclude that the monetary incentive also operated through psychological pathways such as heightened task engagement or issue salience independent of browsing, a limitation discussed below. From the perspective of the literature on information avoidance and disengagement (Toff & Kalogeropoulos, 2020; Wojcieszak et al., 2022), our design does not eliminate disengagement, as we see control-group take-up is very low, but among the participants who do engage, the trajectory appears to lead to attitude change via, in this case, an extrinsic motivational encouragement. This interpretation aligns with recent reasoning by (Tappin et al., 2020), who call for designs that randomize motivation rather than just information; our randomization of motivational encouragements, paired with the IV identification of the effect among compliers, provides one way to better understand these effects.

We also found a directional positive shift in attitudes for child support but no clear directional pattern for cannabis legalization, where attitude movement was distributed across the support-opposition spectrum, with no evidence for polarization. This observation is also consistent with recent updates to motivated-reasoning theory, where deliberation can reduce belief in false news without amplifying polarization (Bago et al., 2020) and information shifts attitudes “in parallel” by similar magnitudes across partisan groups (Coppock et al., 2020). This might be because the motivational frame was epistemic rather than identity-confronting.

**Exploring topic heterogeneity**

The effect of the incentive in our results differs across policies, with significant attitude change for child support and cannabis legalization, but not for the energy transition policy. We offer a post hoc account for this pattern, which is theoretically motivated but should be treated as exploratory given that the content actually consumed by participants was not systematically coded. Two accounts rest on prior literature and are supported by descriptive evidence from our data. First, attitude crystallization: theoretical accounts of attitude strength and pretreatment predict that persuasion effects shrink as priors crystallize through prior exposure through repeated exposure, affective tagging, and partisan cueing (Krosnick & Petty, 1995; Lodge & Taber, 2013), a pattern first experimentally tested by Druckman & Leeper (2012). Public opinion on the energy transition, including debates on solar, wind and especially nuclear power, has been actively contested for over two decades, structured by ideology, regional infrastructure conflict, and identity-relevant cultural meanings (Sonnberger et al., 2021; Sonnberger & Ruddat, 2017). Our exploratory descriptive party-alignment analysis is consistent with this account: participants with far-right or far-left alignments held more extreme attitudes toward energy transition policy at baseline than toward the other two policies. Cannabis legalization and child support policy were by contrast more novel, recent deliberations in parliament and the news, characterized by greater public uncertainty, making information engagement more likely to register. In particular child support also less ideologically pre-loaded, but with much higher cue clarity, as it concerned a concrete administrative reform (a dedicated office and app for streamlined child-support payment); and with German cannabis attitudes strongly tied to expectations about consequences and therefore plausibly sensitive to new information (Manthey et al., 2024).

Second, the information environments compliers encountered differed across topics: fewer participants visited news sources for the energy transition than for child support or cannabis, while energy-transition compliers disproportionately landed on government pages. Whether these source-type differences contributed to the null result for energy transition, for instance through differences in content framing or argumentative quality of the sources, cannot be determined from visit data alone. Additionally, in our data, Google drove much information selection, and the value of self-directed search has been found to depend on what search engines surface in response (Aslett et al., 2024). So, if compliers received via Google an information diet skewed toward governmental sources, this information channel might not have delivered the required attitude-relevant cues.

We hypothesize that two mechanisms could be operating simultaneously: pre-crystallized attitudes on the energy transition may leave less room for revision, and the information channels through which compliers chose to inform themselves might have been less likely to deliver attitude-relevant framing in the first place.

**From abstract policies to specific issues**

We queried attitudes toward specific policy issues, expecting that engagement would produce parallel changes there. The pattern we observed is more selective: multiple significant directed-change effects in the child support policy, and, liable to correction, an effect on the gas/oil heater ban (energy transition) and the cannabis purchase limit. On one hand, the malleability of attitudes is claimed to depend on the granularity of the target, via so-called hierarchical attitude structures, with attitudes towards concrete policy at the bottom below policy domains, with ideological values on top (Peffley & Hurwitz, 1985), with the idea that lower-level attitudes might be easier to affect than “core” political attitudes such as partisan affiliation. On the other hand, through a dual-process lens, specific issues can also be highly technical, and gaining a deeper understanding about local production limits for cannabis, or public land usage for wind and solar parks, may require an understanding of administrative or infrastructural constraints that the typical short search session is unlikely to provide. It is also likely quite difficult to find elite or partisan cues on such specific issues that would guide attitude change via a peripheral route (Gilens & Murakawa, 2002). The items on which we find some evidence of change appear somewhat more salient and less technical (e.g., a dedicated government office and app for child-support payments, a straightforward ban on oil furnaces). This could suggest that when the cognitive cost of processing a piece of information rises, motivation has to do more work to overcome it, and the share of compliers for whom that motivation is sufficient declines. A very low median in search behavior in our data suggests a quite limited dose of information engagement, which is itself informative about the boundary conditions of the mechanism. Further research should investigate the hierarchical nature of attitude malleability further.

**Search-engine centrality and the limits of laboratory analogues**

Our descriptive analysis of compliers' browsing patterns provides a direct illustration of why high-choice environments are difficult to simulate in the laboratory. One, we observe a long tail of websites visited by single participants (exclusive treatments) alongside a small number of URLs that attract many visitors (one cannabis URL was visited by 41% of compliers in that wave, significantly more concentrated than the most-visited URL for the other two policies). Further, participants used commercial search engines as gateways: roughly half of policy-relevant content visits were preceded by a web search, and Google alone concentrated over 80% of all searches. This level of gateway centrality is consistent with what web-tracking work has documented for online news access more broadly (Mangold et al., 2024; Ulloa & Kacperski, 2023): thus, it is possible that the algorithmic ranking of a single company plays a primary role in determining the "informational treatment" individuals actually receive when they choose to inform themselves. Finally, the resulting information environments differ qualitatively across topics: government-heavy for the energy transition, news-heavy for child support and cannabis, in ways that any laboratory analogue would have to specify in advance. Without search engines acting as centralizing forces, it is unlikely that many participants would have converged on any common set of sources. Our results therefore offer empirical support for a theoretical point made by recent work on platform infrastructure (Aslett et al., 2024) that search architecture is itself a treatment that future high-choice environment experiments should consider.

**Limitations**

Several limitations qualify our conclusions. First, the directive to inform oneself may have shaped browsing in ways that diverge from purely incidental exposure. We see this as a boundary condition that may limit but not eliminate external validity: in everyday life, individuals are routinely prompted to seek information by social interaction, curiosity, or professional and educational requirements. Crucially, participants could opt out of the treatment, which limits the risk of reactance from forced exposure. We also designed financial incentives to mirror everyday motivational contexts (e.g., social competition, academic testing) rather than to reward the production of source links, with the aim of minimizing normative influences on policy-support responses. Because the reward is non-normative with respect to attitudinal direction, we do not expect it to bias the attitude items toward demand characteristics; if anything, motivated knowledge-gain is aligned with epistemic engagement rather than with any specific attitudinal direction (Petty & Cacioppo, 1986).

Second, while our behavioral measures such as search counts, URL visits, and time on page allow us to verify that motivational encouragements increased information-seeking, they do not directly index cognitive elaboration, counterarguing, or integrative processing as theorized by dual-process models. The design thus provides evidence that motivation increases engagement with information and that this engagement is associated with attitude change; it does not demonstrate the specific processing pathway. Future studies combining behavioral trace data with process measures (e.g., think-aloud protocols, response-latency indicators, or comprehension checks) would allow this mechanism to be probed more directly. The mindset literature (Ludwig & Sommer, 2024) offers another avenue: experimentally inducing deliberative versus implemental mindsets prior to self-directed search could test whether processing quality, not just engagement quantity, moderates attitude change downstream.

Third, the exclusion restriction underlying our IV strategy, i.e., that encouragement affects attitude change only through verified online information engagement, merits explicit discussion. The monetary incentive is psychologically richer than a behaviorally minimal instrument: it could plausibly heighten task seriousness, increase issue salience, or prompt independent reflection independent of any actual browsing. We cannot fully rule out these alternative pathways. However, several features of the design limit their plausibility: attitude measurement occurs 20 hours after randomization, by which point any salience effect from reading the encouragement text would have substantially dissipated relative to the impact of actual content encountered during browsing, and the knowledge test was at the very end of the survey; our web-tracking data allows us to condition on verified visits, meaning we are not simply comparing self-reports of engagement; and most directly, the verbal encouragement condition, which contained identical task-framing language but no financial stake, produced no significant attitude changes in any wave, despite being equally capable of raising task seriousness or issue salience through its framing. We nonetheless encourage readers to interpret our findings via both the ITT and LATE estimates, acknowledging that residual confounding cannot be fully excluded.

Fourth, web-tracking studies that rely on browser-extension installation share well-known limitations (Makhortykh et al., 2022). Self-selection into the panel constrains generalizability, and awareness of being tracked could in principle have shaped behavior. We note here that participants consented to tracking at least four months prior to the experiment, and many had been part of the panel maintained by our provider for substantially longer (the panel itself is six years old). Long-term participants are plausibly habituated to monitoring, in line with broader evidence on the routine acceptance of commercial tracking (Barth & de Jong, 2017). Additionally, our German sample is a convenience panel and skews male, slightly older, and somewhat more educated than the general population; while the demographic skew is moderate, replication in other panel structures and other political contexts would strengthen the generality of the joint motivation–agency claim.

Finally, our intervention is a short-window encouragement, with attitude measurement 20 hours after baseline. We cannot speak to whether the observed shifts persist over weeks or months, nor to whether repeated motivational activation produces cumulative change. Recent work on the durability of attitude updating through engagement (Costello et al., 2024) on AI-mediated dialogue reducing conspiracy beliefs over months, suggests that durable shifts are achievable with sufficiently engaging informational content, but our design does not test this.

**Conclusion**

This study demonstrates that online information engagement in a high-choice environment can significantly influence policy attitudes when motivation and agency are jointly activated. By embedding randomized motivational encouragement within a relatively unconstrained browsing environment and observing actual web behavior, we provide evidence that selective, motivationally activated online-information engagement can shift policy support, a pattern that theorizes specific conditions (motivational salience, issue malleability, and search-environment quality) under which effects become detectable even in high-choice environments.